\documentstyle[aaspptwo]{article}
\tighten          %single space
\newcommand{\beq}{\begin{equation}}
\newcommand{\eeq}{\end{equation}}
\newcommand{\beqa}{\begin{eqnarray}}
\newcommand{\eeqa}{\end{eqnarray}}
\newcommand{\msun}{\hbox {$M_{\odot}$ }}
\newcommand{\msunt}{\hbox {$M_{\odot}$}}
\newcommand{\dv}{$\Delta {\rm V} ({\rm TO} - {\rm HB})$ }

\newcommand{\dvtwo}{\Delta {\rm V}}

\newcommand{\ea}{{\it et al. }}
\newcommand{\feh}{\hbox{$[{\rm Fe}/{\rm H}]$}}
\newcommand{\dh}{Debye-H\"{u}ckel}
\newcommand{\mvto}{\hbox{$\rm M_v(TO)$}}

\begin{document}

\title{The OPAL Equation of State and Low Metallicity Isochrones}

\author{Brian Chaboyer}

\affil{Canadian Institute for Theoretical Astrophysics, \\
60 St. George Street,  Toronto, Ontario, Canada  M5S 1A7\\
E-Mail: chaboyer@cita.utoronto.ca}

\author{Yong --Cheol Kim}

\affil{Department of Astronomy, and Center for Solar and Space Research\\
Yale University, Box 208101\\
New Haven, CT 06520--8101\\
E-Mail: kim@astro.yale.edu}

\begin{abstract}

The Yale stellar evolution code has been modified to use the OPAL
equation of state tables (Rogers 1994). Stellar models and isochrones
were constructed for low metallicity systems ($-2.8 \le \feh \le
-0.6$).  Above $M\sim 0.7\,\msun$, the isochrones are very similar to
those which are constructed using an equation of state which includes
the analytical \dh ~correction at high temperatures.  The absolute
magnitude of the main sequence turn-off (\mvto ) with the OPAL or \dh
{}~isochrones is about 0.06 magnitudes fainter, at a given age, than
\mvto ~derived from isochrones which do not include the \dh
{}~correction.  As a consequence, globular clusters ages derived using
\mvto ~are reduced by 6 -- 7\% as compared to the ages determined
from the standard isochrones.  Below $M\sim 0.7\,\msun$, the OPAL isochrones
are systematically hotter (by approximately 0.04 in B--V) at a given
magnitude as compared to the standard, or \dh ~isochrones.  However,
the lower mass models fall out of the OPAL table range, and this could
be the cause of the differences in the location of the lower main-sequences.

\end{abstract}


\section{Introduction}

   Low heavy element abundances and a spherical distribution about the
galactic centre are clear indicators that globular clusters were among
the first objects formed in our galaxy.  Consequently it is important
to determine the ages of globular clusters to gain an understanding of
how our galaxy formed and to provide an estimate for the age of the
universe.  The ages of globular clusters are derived by comparing
stellar evolution models to observations of globular cluster colour
magnitude diagrams.  Stellar models are constructed by solving the
basic stellar structure equations.  The solution of the stellar
structure equations require that the opacity, nuclear reaction rates
and equation of state (hereafter EOS) be specified.  Stellar evolution
models are a good representation of the true evolution of a star only
if the input physics used in the solution of the stellar structure
equations are accurate.  As new calculations for the opacity, reaction
rates or EOS become available, it is important to assess the impact of
these improvements on stellar models.  Iglesias \& Rogers (1991)
calculated a new set of opacities which were considerably different
(in certain temperature-density regimes) from previous calculations.
These differences helped to solve many long standing problems in
stellar astrophysics.  The new opacities did not appreciably alter the
ages of the globular clusters, which are best derived using the
absolute magnitude of the main sequence turn-off (\mvto , Chaboyer
1995).  Recently Rogers (1994) has made available EOS tables
appropriate for stellar conditions.  This is the same EOS (hereafter
the OPAL EOS) which was used in the construction of the Iglesias \&
Rogers (1991) opacities.  In order to investigate the effects on the
OPAL EOS on age estimates for globular clusters, the Yale stellar
evolution code has been modified to incorporate the OPAL EOS, and low
metallicity stellar models and isochrones have been constructed.

The importance of the EOS for stellar evolution calculations has long
been recognized.  Eggelton, Faulkner \& Flannery (1973) developed an
EOS that has been widely used in stellar evolution calculations, though
not in the Yale code.  At high temperatures (above $10^6$ K) the
standard EOS in the Yale stellar evolution code is the ideal gas law,
with radiation pressure and degeneracy effects added in the standard
manner (eg.~Schwarzschild 1958).  At low temperatures a Saha equation
(Saha 1920) which includes the single ionization of hydrogen, the
first ionization of the metals and both ionizations of helium is used
as the EOS. This EOS results in typical globular clusters ages of
15 Gyr.  A modification of this standard EOS  includes the effect
of Coulomb forces by using the \dh ~correction. Guenther \ea (1992)
describe the implementation of the \dh ~correction into the Yale
stellar evolution code. The analytical \dh ~correction implemented in
the Yale code is only valid for a fully ionized plasma, so the Saha
equation is still used below $10^6$ K.  Isochrones constructed using
the \dh ~correction yield globular cluster ages which are
approximately 1 Gyr younger than the standard isochrones.  However,
the \dh ~correction is only valid when
\beq \Lambda \equiv
\frac{\sqrt{4\pi} Z^3 e^3 \rho^{1/2}}{(kT)^{3/2}} < 0.2
\eeq
(see Rogers 1994 for a full discussion).  This criterion is often violated
within a stellar plasma.  Thus, it is not clear if including the
\dh ~correction actually results in an improved EOS.

The OPAL EOS includes corrections for Coulomb forces which are
valid over the entire range of temperatures and densities
encountered in standard stellar models, and so
represents a considerable improvement over the analytical \dh
{}~correction.  In \S 2 the construction and evolution of the stellar models is
described, while \S 3 discusses the resulting isochrones.  The
implications that these isochrones have for globular cluster ages is
described  in \S 4.  A brief summary is presented in \S 5.

\section{Stellar Models}

The OPAL EOS tables were integrated into the Yale stellar evolution
code using the interpolation routines supplied by Rogers.  Stellar
models (and isochrones) were then constructed using 3 different EOS:
(1) the standard EOS described in the introduction; (2) the standard
EOS with the analytical \dh ~correction at high temperatures, and (3)
the OPAL EOS.  For each EOS, a series of stellar models with masses
ranging from $M=0.5~\msun$ to $\sim 1.0~\msun$ (in $0.05\,\msun$
increments) were evolved from the zero-age main sequence to the
sub-giant branch.  Numerical difficulties related to the OPAL EOS
prevented us from evolving models on the giant branch.  The numerical
tolerances, and input physics (aside from the EOS) were identical for
all evolutionary runs.  In order to span the range of metallicities
encountered in the globular cluster system, models were evolved with
$Z = 6\times 10^{-5}$, $2\times 10^{-4}$, $6\times 10^{-4}$, $2\times
10^{-3}$, $4\times 10^{-3}$, and $7\times 10^{-3}$, corresponding to
$\feh = -2.8; ~ -2.3; ~-1.8; ~-1.3$ all with $[\alpha/{\rm Fe}] =
+0.40$; $\feh = -0.9$, $[\alpha/{\rm Fe}] = +0.30$; and $\feh = -0.6$,
$[\alpha/{\rm Fe}] = +0.20$. The models use a scaled solar
composition, and the effect of $\alpha$-element enhancement has been
taken into account by modifying the relationship between $Z$ and \feh,
as prescribed by Salaris, Chieffi \& Straniero (1993).

All models included the effects of $^4$He diffusion, using the
diffusion coefficients of Michaud \& Proffitt (1993).  A mixing length
of $\alpha = 1.7$ was used in all of the models. The various EOS
require mixing lengths which differ by less than 2\% from each other
for a solar calibrated model, so it was not deemed necessary to use
different mixing lengths in the low metallicity models.  The other
input physics used in the models are the standard ones for the Yale
code, as described by Chaboyer (1995).

The effect of the OPAL EOS on the structure of the stellar models is
illustrated in Figure 1a, which compares the run of temperature and
density for $M = 0.5$ and $0.8\,\msun$ stellar models using the OPAL
and \dh ~EOS.  The differences between the two EOS are small.
Unfortunately, the $M= 0.5\,\msun$ model falls outside the OPAL EOS
around $\log T = 5$.  When the models fall outside the OPAL tables,
the standard EOS has been substituted.  Figure 1b plots the difference
in density ($\Delta\rho/\rho$) as a function of temperature, where the
pressure and temperature have been taken from the \dh ~models.  We see
that for a given temperature and pressure the OPAL EOS gives densities
which typically differ by $2 - 4\%$ from the \dh ~case.  This plot also
makes it clear where the $M = 0.5\,\msun$ model falls outside the
table, for these regions are identified by the solid line at
$\Delta\rho/\rho = 0$ in Figure 1b.  We have made similar plots for
other masses, and stages of evolution, and have determined that all
models with $M\ga 0.75\,\msun$ stay within the OPAL EOS tables
throughout their evolution.  Lower mass models fall outside the tables
near the zero-age main sequence (ZAMS), but are usually entirely within
the OPAL tables as they evolve off of the ZAMS.  Unfortunately, for
low mass models ($M\la 0.7\,\msun$), it is the ZAMS models which are
most important, as such low mass stars are still essentially
unevolved, even in globular clusters.  For this reason, we must view
with caution properties of the OPAL EOS models and isochrones below $M
\simeq 0.7\,\msun$, for they may be affected by the fact that the EOS
abruptly changes to the standard one in part of the model.

The effects of the \dh ~and OPAL EOS on stellar evolutionary tracks is
shown in Figure 2 ($M = 0.6\,\msun$) and Figure 3 ($M = 0.8\,\msun$).
Figure 2 demonstrates that the low mass OPAL models are shifted to
hotter effective temperatures near the ZAMS as compared to the \dh
{}~models.  By the time the models reach the turn-off (and the model is
entirely within the OPAL table), the OPAL and \dh ~evolutionary tracks
are nearly identical.  This suggests that the shift near the ZAMS may
be due to the fact that the model is falling outside the OPAL table
around $\log T = 5$.  Figure 3 shows that for higher mass models, the
evolution is nearly identical for the OPAL and \dh ~cases.  Thus, the
turn-off properties of isochrones constructed from the OPAL or \dh
{}~models should be similar.  In order to demonstrate the possible
effect that falling out of the OPAL tables has on the low mass models,
Figure 3 plots a model in which the OPAL EOS has been used except for
$\log T = 4.4 - 5.2$, in which case the standard EOS was substituted.
Note that this model is shifted in cooler effective temperature as
compared to the pure OPAL, or \dh ~cases.  This shift is in the
opposite direction than that which occurs in the low mass models.
However, it does demonstrate that evolutionary tracks can be
substantially modified if the OPAL EOS is not used in the entire
model.  Thus, the shift in effective temperature found in the low mass
OPAL models may be due to the fact that these models fall outside the
OPAL EOS tables around $\log T = 5$.

\section{Isochrones}

Isochrones were constructed by interpolating  the evolutionary
tracks using the method of equal evolutionary points (Prather 1976)
for the different $Z$ values and EOS discussed above.  The isochrones
spanned the age range 8 -- 22 Gyr, in 1 Gyr increments. The colour
transformation of Green, Demarque \& King (1987) was used to transform
from the theoretical luminosities and temperatures to the UBVRI
system.  Figure 4 plots the 14 Gyr, $\feh = -1.3$ isochrones for the
different EOS. Brighter than $M_V \simeq 5.2$, the \dh ~and OPAL
isochrones are very similar.  They are somewhat redder, and have a
fainter main sequence turn-off magnitude as compared to the standard
isochrones.  Fainter than $M_V \simeq 5.2$, the standard and \dh
{}~isochrones are very similar.  The OPAL isochrones are somewhat
hotter (at a given magnitude) as compared to the other isochrones along
the lower main sequence.  The above statements are true for all of the
isochrones we have constructed.  The isochrones are always quite
similar around $M_V = 5.2$.  This corresponds to a mass of roughly
$0.7\msun$.  This critical mass is a function of the metallicity and
age of the isochrone, and the varies between $0.67$ and $0.74\,\msun$.
This is a similar mass range to where the stellar models fall out of
the OPAL EOS tables (around $\log T = 5$).  As discussed in the
previous section, the change in location of the lower main sequence in
the OPAL isochrones (as compared to the standard or \dh ~isochrones)
is likely due to the fact that we have been forced to use the standard
EOS around $\log T = 5$ for the lower mass models.

In order to determine which set of isochrones most closely match the
observations of globular clusters, we have fit the isochrones to the
fiducial sequences of M92 (Stetson \& Harris 1988), and NGC 288 (Bolte
1992).  The fit to M92 (with $\feh = -2.3$) is shown in Figure 5.  The
best fit to the data was found using the objective near point
estimator technique of Flannery \& Johnson (1982; see also Durrell \&
Harris 1993).  In performing this fit, the distance modulus was
allowed to vary by $\pm 0.1$ mag from its nominal value of $14.52$,
and the reddening to vary between 0.0 and 0.05 (its nominal value is
0.02).  For all three sets of isochrones, the best fit was found with
an age of 17 Gyr.  The standard isochrones are best able to reproduce
the shape of the observed CMD.  The \dh ~isochrones (not shown)
produced a fit which was nearly as good.  The OPAL isochrones do not
fit the data as well;  this is due to the fact that the slope of the
lower main sequence ($\Delta V/\Delta (B-V)$) is too steep in the OPAL
isochrones.  Although not shown here, very similar results are found
for the fit to NGC 288.

The fact that the OPAL isochrones do not fit the observations as well
as the standard, or \dh ~isochrones does not imply that the OPAL EOS
is inferior to the standard or \dh ~EOS.  The shape of the isochrones
is influenced by our colour calibration, choice of mixing length,
surface boundary conditions, as well as the fact that the
lower mass models fall out of the OPAL EOS tables.  All that may be
concluded from Figure 5 is that for our choice of input physics, the
standard isochrones are best able to reproduce the morphology of
observed colour-magnitude diagrams.

\section{Globular Cluster Ages}
Fitting of isochrones to observed colour magnitude diagrams is subject
to numerous theoretical uncertainties (colour calibration, mixing
length theory, model atmospheres, etc).  For this reason, it is best
to determine the ages of globular clusters using the absolute
magnitude of the main sequence turn-off (\mvto).  This luminosity
based age indicator is subject to far fewer theoretical uncertainties
than age indicators which use colour information.  For each set of
isochrones, \mvto ~was determined by taking the average magnitude
of the bluest point on the isochrones.   For each metallicity, the age
in Gyr ($t_9$) was fit as a function of \mvto ~ using
an equation of the form:
\beq
t_9 = \beta_0 + \beta_1 \mvto + \beta_2\mvto^2.
\eeq
This simple quadratic fit did an excellent job of reproducing the
mean trend in the calculated \mvto, while removing some of the
`wiggles' introduced by the isochrone construction program.  Table 1
give the coefficients of the fit for the standard and OPAL isochrones.  As
Figure 4 demonstrates, \mvto ~for the \dh ~and OPAL isochrones are very
similar, so the fit coefficients for the \dh ~isochrones have not been
included in Table 1.
\begin{planotable}{lrrrrrrr}
\tablecaption{TABLE 1}
\tablecaption{Modeling Coefficients}
\tablehead{
& \multicolumn{3}{c}{Standard} &&
\multicolumn{3}{c}{OPAL}\nl
\cline{2-4}\cline{6-8}
{}~\\[-8pt]
\colhead{\feh}&
\colhead{$\beta_1$} &
\colhead{$\beta_2$} &
\colhead{$\beta_3$} & &
\colhead{$\beta_1$} &
\colhead{$\beta_2$} &
\colhead{$\beta_3$}
}
\startdata
$-2.82$ & 60.595 & $-38.353$ & 6.945 &&
$  61.079$ & $ -37.933$ & $6.730$\nl
$-2.29$ &$  72.007$ & $ -43.525$ & $   7.319$&&
$  85.839$ & $ -50.317$ & $   8.088$\nl
$-1.82$ &$  73.431$ & $ -43.722$ & $   7.160$&&
$  79.331$ & $ -45.804$ & $   7.252$\nl
$-1.29$ &$ 101.546$ & $ -56.583$ & $   8.469$&&
$  84.699$ & $ -47.586$ & $   7.240$\nl
$-0.91$ &$  96.273$ & $ -53.127$ & $   7.873$&&
$  94.421$ & $ -51.422$ & $   7.532$\nl
$-0.59$ &$  93.204$ & $ -51.140$ & $   7.531$&&
$  89.324$ & $ -48.428$ & $   7.071$\nl
\end{planotable}

Figure 6 plots the fit of age as a function of \mvto ~for the
$\feh = -1.3$ isochrones.  This plot demonstrates the similarity in
\mvto ~for the \dh ~and OPAL isochrones.  For a given value of
\mvto ~the standard isochrones imply ages that are approximately
$0.5 - 1.5$ Gyr older than the OPAL or \dh ~isochrones.  It is
worthwhile noting that the
greatest age reduction occurs for the oldest isochrones.  The trends
seen here are true for all metallicities that have been
calculated. Thus, globular cluster ages determined using \mvto ~will
be approximately 1 Gyr younger with the \dh ~or OPAL isochrones,
as compared to the standard isochrones.

In order to apply our isochrones to actual data, it is necessary
to convert observed
magnitudes into absolute magnitudes.  This may be accomplished in an
reddening independent way by comparing the difference between \mvto
{}~and the magnitude of the horizontal branch (in the RR Lyr instability
strip).  This age determination technique is referred to as \dv.  In
order to use \dv ~to determine the ages, we must specify the absolute
magnitude of the RR Lyr stars ($M_V ({\rm RR})$).  There are a number
of different ways to determine $M_V ({\rm RR})$, all of which find
that $M_V ({\rm RR}) = \mu\, \feh + \gamma$
where $\mu$ is the slope with metallicity, and $\gamma$ is the
zero-point.  Estimates for the slope vary from 0.15 to 0.30.  For this
study, a value of $\mu = 0.20$ has been chosen, which is in reasonable
agreement with Baade-Wesslink and infrared flux methods for field RR
Lyr stars (Carney, Storm \& Jones 1992; Skillen \ea 1993), and
theoretical calculations (Lee 1990).  For the zero-point, a value of
$\gamma = 0.94$ has been chosen.  This value is simply the average of
the values determined by Walker (1992) who observed LMC RR Lyrs
(using the SN1987A distance to the LMC) and Layden \ea (1995) who
determined $M_V ({\rm RR})$ using statistical parallax observations.
Thus, the following formulae was used to determine the absolute
magnitude of the RR Lyr stars:
\beq
M_V ({\rm RR}) = 0.20\, \feh + 0.94.
\label{mvrr}
\eeq

The \mvto ~values determined from the isochrones
are combined with equation (3) to determine \dv as a function
of metallicity and age (in Gyr).  The
resulting grid was modeled using an equation of the form
\beqa
t_9 & = & a_o + a_1\dvtwo + a_2\dvtwo ^2 + a_3\feh \nonumber \\
&+& a_4\feh ^2 +a_5\dvtwo \feh
\label{age}
\eeqa
and globular cluster ages were determined using the above equation for
each EOS.  Table 2 lists the ages for 40 globular clusters with
observed \dv ~values which are within the metallicity range of our set
of isochrones.  The observed \dv ~are taken primarily from the
compilation of Chaboyer, Sarajedini \& Demarque (1992).  A few
measurements are from the
compilation of Walker (1992).  The observed \dv for NGC 6535 is from
Sarajedini (1994), and that of NGC 6652 is from Ortolani, Bica \&
Barbuy (1994).  The cluster metallicities are taken primarily from
Zinn \& West (1984), with a few exceptions as noted by Chaboyer \ea
(1992). The error in the derived age was determined by propagating
through the errors in \dv and \feh ~as determined by the observers. The
average age of the globular clusters is 13.90 Gyr for the standard
isochrones, 13.03 Gyr for the \dh ~isochrones and 12.97 Gyr for the
OPAL isochrones.  Thus, as expected, the OPAL EOS results in ages
similar to the \dh ~EOS, both of which about 0.9 Gyr younger than the
standard isochrones.  Note that there is evidence for an age range
within the globular clusters, so the average ages quoted above should
not be taken  as `the' age of the globular clusters.  The
difference in age as determined by the OPAL and standard isochrones is
plotted as a function of metallicity and age in Figure 7.  It is clear
that the largest age reductions occur for the oldest clusters.  The
age reduction is not as well correlated with metallicity, although the
largest age reductions tend to occur for the most metal-poor clusters;
this is likely due to the fact that the metal-poor clusters also tend
to be the oldest clusters.  As the zero-point of our ages is set by
the zero-point in the $M_V ({\rm RR})$ relation (eq. 3),
larger age reductions will occur if a brighter zero-point for $M_V
({\rm RR})$ is adopted.
\begin{planotable}{rlrrrrr}
\tablecaption{TABLE 2}
\tablecaption{Globular Cluster Ages}
\tablehead{
&&&&
\colhead{Standard }&
\colhead{\dh }&
\colhead{OPAL}\nl
\colhead{NGC}&
\colhead{Name}&
\colhead{\feh}&
\colhead{$\dvtwo$}&
\colhead{Age (Gyr)}&
\colhead{Age (Gyr)}&
\colhead{Age (Gyr)}
}
\startdata
 104& 47 Tuc& $-0.71\pm 0.08$ & $3.61\pm 0.10$ & $14.29\pm 1.9$& $13.20\pm
1.7$& $13.23\pm 1.7$\nl
 288&       & $-1.40\pm 0.12$ & $3.60\pm 0.12$ & $14.82\pm 1.9$& $13.91\pm
1.7$& $13.83\pm 1.7$\nl
 362&       & $-1.27\pm 0.07$ & $3.42\pm 0.14$ & $12.41\pm 2.4$& $11.59\pm
2.2$& $11.55\pm 2.2$\nl
1261&       & $-1.31\pm 0.09$ & $3.44\pm 0.12$ & $12.40\pm 1.6$& $11.68\pm
1.5$& $11.62\pm 1.5$\nl
1851&       & $-1.36\pm 0.09$ & $3.45\pm 0.10$ & $12.58\pm 1.4$& $11.85\pm
1.2$& $11.79\pm 1.2$\nl
1904& M79   & $-1.69\pm 0.09$ & $3.45\pm 0.12$ & $13.01\pm 1.7$& $12.23\pm
1.5$& $12.16\pm 1.5$\nl
2298&       & $-1.85\pm 0.11$ & $3.49\pm 0.21$ & $13.84\pm 3.0$& $12.97\pm
2.8$& $12.90\pm 2.8$\nl
2808&       & $-1.37\pm 0.09$ & $3.50\pm 0.14$ & $13.29\pm 2.0$& $12.50\pm
1.8$& $12.43\pm 1.8$\nl
3201&       & $-1.61\pm 0.12$ & $3.39\pm 0.17$ & $12.10\pm 2.2$& $11.40\pm
2.0$& $11.34\pm 2.0$\nl
4147&       & $-1.80\pm 0.26$ & $3.60\pm 0.12$ & $15.43\pm 2.0$& $14.44\pm
1.8$& $14.36\pm 1.8$\nl
4590& M68   & $-2.09\pm 0.11$ & $3.42\pm 0.10$ & $13.30\pm 1.4$& $12.47\pm
1.3$& $12.41\pm 1.3$\nl
5024& M53   & $-2.04\pm 0.08$ & $3.56\pm 0.14$ & $15.24\pm 2.2$& $14.26\pm
2.0$& $14.19\pm 2.0$\nl
5053&       & $-2.41\pm 0.06$ & $3.48\pm 0.12$ & $14.84\pm 1.8$& $13.88\pm
1.6$& $13.83\pm 1.6$\nl
5272& M3    & $-1.66\pm 0.06$ & $3.54\pm 0.09$ & $14.26\pm 1.4$& $13.38\pm
1.2$& $13.30\pm 1.2$\nl
5466&       & $-2.22\pm 0.36$ & $3.58\pm 0.12$ & $15.94\pm 2.1$& $14.89\pm
1.9$& $14.83\pm 1.9$\nl
5897&       & $-1.68\pm 0.11$ & $3.60\pm 0.18$ & $15.23\pm 2.9$& $14.26\pm
2.6$& $14.18\pm 2.6$\nl
5904& M5    & $-1.40\pm 0.06$ & $3.49\pm 0.11$ & $13.18\pm 1.6$& $12.40\pm
1.4$& $12.33\pm 1.4$\nl
6101&       & $-1.81\pm 0.15$ & $3.40\pm 0.12$ & $12.53\pm 1.6$& $11.78\pm
1.5$& $11.72\pm 1.4$\nl
6121& M4    & $-1.33\pm 0.10$ & $3.55\pm 0.16$ & $13.97\pm 2.4$& $13.13\pm
2.2$& $13.06\pm 2.2$\nl
6171& M107  & $-0.99\pm 0.06$ & $3.60\pm 0.18$ & $14.40\pm 2.8$& $13.55\pm
2.6$& $13.50\pm 2.5$\nl
6205& M13   & $-1.65\pm 0.06$ & $3.55\pm 0.21$ & $14.40\pm 3.2$& $13.50\pm
2.9$& $13.43\pm 2.9$\nl
6218& M12   & $-1.34\pm 0.09$ & $3.45\pm 0.14$ & $12.56\pm 1.9$& $11.83\pm
1.7$& $11.77\pm 1.7$\nl
\phantom{0}6254& M10   & $-1.75\pm 0.08$ & $3.75\pm 0.15$ & $17.92\pm 2.7$&
$16.73\pm 2.5$& $16.65\pm 2.5$\nl
6341& M92   & $-2.24\pm 0.08$ & $3.65\pm 0.12$ & $17.15\pm 2.1$& $16.00\pm
1.9$& $15.94\pm 1.9$\nl
6397&       & $-1.91\pm 0.14$ & $3.64\pm 0.14$ & $16.29\pm 2.4$& $15.22\pm
2.2$& $15.15\pm 2.2$\nl
6584&       & $-1.54\pm 0.15$ & $3.47\pm 0.12$ & $13.08\pm 1.7$& $12.29\pm
1.5$& $12.23\pm 1.5$\nl
6535&       & $-1.75\pm 0.15$ & $3.66\pm 0.19$ & $16.33\pm 3.2$& $15.27\pm
3.0$& $15.19\pm 3.0$\nl
6652&       & $-0.89\pm 0.15$ & $3.35\pm 0.16$ & $10.28\pm 2.3$& $ 9.52\pm
2.1$& $ 9.53\pm 2.1$\nl
6752&       & $-1.54\pm 0.09$ & $3.65\pm 0.16$ & $15.82\pm 2.7$& $14.82\pm
2.4$& $14.74\pm 2.4$\nl
6809& M55   & $-1.82\pm 0.15$ & $3.55\pm 0.10$ & $14.68\pm 1.6$& $13.75\pm
1.4$& $13.67\pm 1.4$\nl
7006&       & $-1.59\pm 0.07$ & $3.55\pm 0.12$ & $14.31\pm 1.8$& $13.42\pm
1.7$& $13.35\pm 1.7$\nl
7078& M15   & $-2.15\pm 0.08$ & $3.54\pm 0.16$ & $15.16\pm 2.5$& $14.18\pm
2.3$& $14.11\pm 2.3$\nl
7099& M30   & $-2.13\pm 0.13$ & $3.53\pm 0.14$ & $14.96\pm 2.2$& $14.00\pm
2.0$& $13.93\pm 2.0$\nl
\phantom{0}7492&       & $-1.82\pm 0.30$ & $3.61\pm 0.14$ & $15.62\pm 2.3$&
$14.62\pm 2.1$& $14.54\pm 2.1$\nl
    & Ter8  & $-1.99\pm 0.08$ & $3.65\pm 0.12$ & $16.61\pm 2.0$& $15.52\pm
1.9$& $15.44\pm 1.9$\nl
    & Rup106& $-1.69\pm 0.05$ & $3.15\pm 0.12$ & $10.20\pm 1.6$& $10.06\pm
1.6$& $ 9.94\pm 1.5$\nl
    & Pal5  & $-1.47\pm 0.29$ & $3.40\pm 0.14$ & $12.05\pm 1.8$& $11.36\pm
1.7$& $11.30\pm 1.7$\nl
    & Pal12 & $-1.14\pm 0.20$ & $3.30\pm 0.12$ & $10.11\pm 1.8$& $ 9.38\pm
1.7$& $ 9.38\pm 1.7$\nl
    & IC4499& $-1.50\pm 0.20$ & $3.25\pm 0.15$ & $10.33\pm 1.6$& $ 9.78\pm
1.5$& $ 9.74\pm 1.4$\nl
    & Arp2  & $-1.70\pm 0.11$ & $3.29\pm 0.10$ & $11.02\pm 1.1$& $10.40\pm
1.0$& $10.35\pm 1.0$\nl
\end{planotable}

\section{Summary}
The OPAL EOS tables from Rogers (1994) have been integrated into the
Yale stellar evolution code.  Stellar models and isochrones were
constructed with the standard EOS, an equation of state which includes
the \dh ~correction at high temperatures, and the OPAL EOS.  These
calculations covered a range in metallicity ($-2.8 \le \feh -0.6$) and
age (8 -- 22 Gyr) appropriate for the study of globular clusters.
Lower mass models (below $M \la 0.7\,\msun$) with the OPAL EOS
are shifted to hotter effective temperatures as compared to the
standard or \dh ~models.  However, the lower mass models fall out of the OPAL
EOS tables around $\log T = 5$.  A test was performed whereby a
$0.8\,\msun$ stellar model (which is normally entirely within the OPAL
EOS tables) was evolved with the OPAL EOS except near $\log T = 5$,
where the standard EOS was substituted.  This evolutionary track was
shifted to cooler effective temperature.  Although the shift was in the
opposite direction from that found in the lower mass models, it nevertheless
points out that the cause of the shift in the H-R diagram for lower
mass stars could be due to the fact that these models fall outside the
OPAL EOS tables. This question can only be answered definitively if
the coverage of OPAL EOS tables is expanded so that low mass stars do
not fall out of the tables.  This is of some importance, as it was
determined that the OPAL isochrones do not match the shape of
observed colour magnitudes diagrams on the lower main sequence,

Above $M \simeq 0.7\,\msun$, the OPAL and Debye-H\"{u}ckel stellar
evolution tracks and isochrones are very similar.  They predict a
main-sequence turn-off magnitude which is fainter at a given age as
compared to the standard isochrones.  Hence, age determinations which
are based on the main sequence turn-off magnitude will be
systematically younger (by 0.5 -- 1.5 Gyr) for the OPAL or \dh
{}~isochrones.  The greatest age reductions occur for the oldest
clusters.  This result has been verified by determining \dv ages for
40 globular clusters (Table 2).  The OPAL and \dh ~isochrones yield
very similar ages, and are 6 -- 7\% lower than the standard
isochrones.

\acknowledgments
We would like to thank F.~Rogers for providing us with his EOS tables
and interpolation software, and answering our many questions.
Research supported in part by NASA grants
NAG5--1486 and  NAGW--2531 to Yale University.

%\clearpage
\appendix

\clearpage

\begin{center}
{\bf Figure Captions}
\end{center}
\begin{description}

\item[Figure 1:]The upper panel, (a), plots the run of temperature and
density for $M = 0.5$ and $0.8\,\msunt$, $Z=2\times 10^4$ stellar
models near the zero-age main sequence.  The solid line is for models
which use the OPAL EOS, while the dashed line is for models which use
the \dh ~EOS.  The lower panel, (b) plots the difference in density,
$\Delta \rho/\rho \equiv [\rho_{\rm Debye-H\ddot{u}ckel} - \rho_{\rm OPAL} ]/
\rho_{\rm Debye-H\ddot{u}ckel}$
for $M=0.5$ and $0.8\,\msunt$ stellar
models.  In this plot, the run of temperature and pressure was taken
from the \dh ~models, so that the differences are solely due to the
different EOS.  The solid line at $\Delta \rho/\rho = 0$ for the
$M = 0.5\,\msunt$ model is due to the fact that the model is outside
the OPAL table around $\log T = 5$, and the \dh ~EOS was substituted.

\item[Figure 2:]The stellar evolution tracks for the $M = 0.6\msunt$,
$Z=2\times 10^4$ models  with the \dh ~EOS and OPAL EOS.
Note that near the turn-off (where the model is entirely within the
OPAL table), the two tracks nearly coincide.

\item[Figure 3:]The stellar evolution tracks for the $M = 0.8\msunt$,
$Z=2\times 10^4$ models with different EOS.  The \dh ~EOS
and OPAL EOS evolutionary tracks are nearly identical.  A model where
the standard EOS was substituted for the OPAL EOS for $\log T = 4.4 -
5.2$ is shown as the short-dashed line to demonstrate the effect that
falling out of the OPAL EOS has on the evolution of the model (this
situation occurs for the lower mass models).

\item[Figure 4:]Isochrones with different EOSs and an age of 14 Gyr
with $\feh = -1.3$ are plotted in the B, B--V plane.

\item[Figure 5:]The fit between the isochrones and the M92 fiducial
sequence (solid line connecting the points) of Stetson \& Harris
(1988).  The dashed lines are the best fitting
isochrones.  The left panel (a) shows the standard
isochrone, while the right panel (b) plots the OPAL isochrone.  The
standard isochrones provide an excellent match to the data, and imply
an age of 17 Gyr.  The OPAL isochrones do not match the data as well.

\item[Figure 6:]Age (in Gyr) as a function of the absolute magnitude
of the main sequence turn-off (\mvto) for the $\feh
= -1.3$ isochrones.  Note that the the OPAL and \dh ~isochrones are
very similar, while the standard isochrones imply ages that are
somewhat older, for a given \mvto.

\item[Figure 7:]The reduction in age implied by the OPAL EOS
($\rm Age_{Standard} - Age_{OPAL}$) is plotted as a function of
metallicity in the left panel (a), and as a function of the
standard age in the right panel, (b).

\end{description}
\end{document}